\documentclass[a4paper,fleqn]{cas-dc}

\usepackage[square,numbers]{natbib}
\usepackage{amsmath,amsfonts}
\usepackage{algorithm,algorithmic}
\usepackage{array}
\usepackage{float}
\usepackage{textcomp}
\usepackage{stfloats}
\usepackage{url}
\usepackage{verbatim}
\usepackage{subcaption}
\usepackage{graphicx}
\usepackage{makecell}
\usepackage{xcolor}
\usepackage{hyperref}
\usepackage{tablefootnote}
\usepackage{scalerel}
\usepackage{amssymb}
\usepackage{pifont}
\usepackage{placeins}

\newcommand{\cmark}{\ding{51} }
\newcommand{\xmark}{\ding{55} }

\newcolumntype{P}[1]{>{\raggedright\arraybackslash}p{#1}}
\newcolumntype{C}[1]{>{\centering\arraybackslash}p{#1}}

\usepackage{etoolbox}
\usepackage{soul} 

\newtoggle{finalPaper}

\setstcolor{red}

\togglefalse{finalPaper} 

\iftoggle{finalPaper} {

	\newcommand{\rmvtxt}[1]{}}
{

	\newcommand{\rmvtxt}[1]{\st{#1}}}

\begin{document}

\let\WriteBookmarks\relax
\def\floatpagepagefraction{1}
\def\textpagefraction{.001}
\shorttitle{Robust Federated Learning for execution time-based device model identification}
\shortauthors{S\'anchez et~al.}

\title[mode=title]{Robust Federated Learning for execution time-based device model identification under label-flipping attack}

\author[1]{Pedro Miguel {S\'anchez S\'anchez}}[orcid=0000-0002-6444-2102]
\cormark[1]

\author[2]{Alberto {Huertas Celdr\'an}}[orcid=0000-0001-7125-1710]

\author[1]{Jos\'e Rafael {Buend\'ia Rubio}}

\author[3]{G\'er\^ome {Bovet}}[orcid=0000-0002-4534-3483]

\author[1]{Gregorio {Mart\'inez P\'erez}}[orcid=0000-0001-5532-6604]

\address[1]{Department of Information and Communications Engineering, University of Murcia, Murcia 30100, Spain}

\address[2]{Communication Systems Group (CSG), Department of Informatics (IfI), University of Zurich UZH, 8050 Zürich, Switzerland}

\address[3]{Cyber-Defence Campus within armasuisse Science \& Technology, CH---3602 Thun, Switzerland}

\cortext[cor1]{Corresponding author.
Email address: pedromiguel.sanchez@um.es (P.M.S. S\'anchez)}


\maketitle

\begin{abstract}
The computing device deployment explosion experienced in recent years, motivated by the advances of technologies such as Internet-of-Things (IoT) and 5G, has led to a global scenario with increasing cybersecurity risks and threats. Among them, device spoofing and impersonation cyberattacks stand out due to their impact and, usually, low complexity required to be launched. To solve this issue, several solutions have emerged to identify device models and types based on the combination of behavioral fingerprinting and Machine/Deep Learning (ML/DL) techniques. However, these solutions are not appropriated for scenarios where data privacy and protection is a must, as they require data centralization for processing. In this context, newer approaches such as Federated Learning (FL) have not been fully explored yet, especially when malicious clients are present in the scenario setup. The present work analyzes and compares the device model identification performance of a centralized DL model with an FL one while using execution time-based events. For experimental purposes, a dataset containing execution-time features of 55 Raspberry Pis belonging to four different models has been collected and published. Using this dataset, the proposed solution achieved 0.9999 accuracy in both setups, centralized and federated, showing no performance decrease while preserving data privacy. Later, the impact of a label-flipping attack during the federated model training is evaluated, using several aggregation mechanisms as countermeasure. Zeno and coordinate-wise median aggregation show the best performance, although their performance greatly degrades when the percentage of fully malicious clients (all training samples poisoned) grows over 50\%.
\end{abstract}



\section{Introduction}

Currently, there exist a vast number of devices deployed all over the world, from smart cars, traffic lights, security systems, to smart homes and industries. The IoT market has grown to a total of 31 billion connected devices by 2020, with a forecast of $\approx$30 billion devices connected to each other by 2023, according to Cisco \cite{cisco_report}. One of the main reasons of this growth is the fourth industrial revolution or Industry 4.0, with the explosion of a set of technologies and paradigms such as 5G, machine and deep learning (ML/DL), robotics, and cloud computing.

The emergence of such technologies poses new challenges to be solved in order to ensure a safe and efficient environment \cite{nivzetic2020internet}. In this sense, there are billions of connected devices, many of them performing critical tasks where failures can be fatal, such as autonomous car driving or industrial operations. In addition, the growing popularity of these technologies makes them a desirable target for cybercriminals. Between the possible security threats affecting resource constrained devices, device impersonation is one of the most serious problems of large organizations with proprietary hardware where one device model could be impersonated for malicious purposes, such as industrial espionage. In addition, there are a multitude of counterfeit devices on the market, some of which are difficult to differentiate from the original \cite{negka2019employing}.

To solve these issues, device model and type identification based on performance fingerprinting arises as a solution \cite{sanchez2020survey}. The main benefit of device model identification is to prevent third-party attacks such as spoofing, as well as to identify malicious or counterfeit devices. Although there are numerous works in the literature exploring the identification of models from different performance characteristics, such as execution-time, network connections or system logs, and leveraging ML/DL for data processing, these solutions mostly require data centralization, making them not suitable for scenarios where data leakage protection and privacy is critical. In this sense, Federated Learning (FL) based techniques have recently gained enormous prominence \cite{yang2019federated}. In FL approaches, the training data of the ML/DL models remain private and while the locally trained models are shared. Later, these models are aggregated (usually by a central party) into a joint model that goes back to the clients for further training, repeating the process in a cyclic fashion. This approach improves both the privacy of the data, as it does not leave the client but also the communication overhead, as sharing only model parameters is usually less resource consuming that sharing the complete data used for training.

In addition, there are few datasets modeling the performance of IoT devices for identification \cite{sanchez2020survey}, and any of them is focused on execution time performance or FL-based scenarios. Moreover, most of the current solutions in the literature do not explore the impact of possible adversarial attacks targeting the ML/DL models during their generation and deployment \cite{Nguyen2019DioT}. These attacks may happen when one of the clients participating in the federation acts in a malicious way sending corrupted model updates. These problems have additional importance in FL setups, where the control of the clients is no longer under the entity generating the joint ML/DL model.

Therefore, this work explores the following three main areas to improve the completeness of the literature: \textit{(i)} the identification of device models using centralized Machine Learning (ML) algorithms and execution time data, \textit{(ii)} the decentralization of this training using the Federated Learning (FL) techniques, and \textit{(iii)} the use of the Adversarial Machine Learning (AML) techniques to evaluate and improve the robustness of the generated models. In this sense, its main contributions are:

\begin{itemize}
    \item An execution time-based performance dataset collected in 55 different Raspberry Pi (RPi) devices from 4 different models, and intended for model identification. This dataset is generated using physical devices under normal functioning, reflecting a real scenario where many devices are operating.
    
    \item The comparison between a centralized and a federated Multi-Layer Perceptron (MLP) model with identical configuration, only changing its training approach. It is showed how the federated setup maintains an almost identical model identification accuracy of 0.9999, without losing performance and improving data leakage protection and privacy.
    
    \item The comparison of different aggregation methods as countermeasure for the federated model under a label-flipping attack. Federated averaging, coordinate-wise median, Krum and Zeno aggregation methods are compared, showing median and Zeno the best results regarding attack resilience.
    
\end{itemize}

The remainder of this paper is structured as follows. Section \ref{sec:related} describes the closest works in the literature, motivating this research. Section \ref{sec:dataset} explains the procedure followed to extract the model identification data. Later, Section \ref{sec:classification} compares the performance of a DL-based classifier when it is trained from a centralized and from a federated approach. Section \ref{sec:adversarial} explains the adversarial setup followed to test the solution resilience against attacks. Finally, Section \ref{sec:conclusions} draws the conclusions extracted from the present research and future lines to explore.

\section{Related Work}
\label{sec:related}

This section will review how the problem of device identification has been addressed to date from different approaches and techniques. Likewise, some works in the literature on Federated Learning and Adversarial Machine Learning will be analyzed.

Device type and model identification has been widely explored in the literature, with varied data sources and ML/DL-based processing techniques \cite{sanchez2020survey}. As one of the closest works to the present one, the authors of \cite{babun2021cps} proposed a novel challenge-response fingerprinting framework called STOP-AND-FRISK (S\&F) to identify classes of Cyber-Physical Systems (CPS) devices and complement traditional CPS security mechanisms based on hardware and OS/kernel. It is exposed that unauthorized and spoofed devices may include manipulated pieces of software or hardware components that may adversely affect CPS operations or collect vital CPS metrics from the network. Another interesting paper showing a fingerprinting technique using hardware performance is \cite{sanchez2018clock}. Such a technique is based on the execution times of instruction sequences available in API functions. Due to its simplicity, this method can also be performed remotely. Additionally, network is the main data source employed in the literature for device model and type identification \cite{meidan2017profiliot}, as it can be collected from an external gateway.

Regarding the application of FL in device identification, the authors of \cite{he2021edge} leveraged FL for device type identification using network-based features. Here, authors experienced a slightly reduced performance compared to a centralized setup, 0.851 F1-score in the centralized and 0.849 in the federated, but the training process was faster and safer. Additionally, in \cite{mun2021internet}, the authors performed application type classification based on network traffic using FL to build the models. Although the authors of \cite{Thangavelu2018DEFT} proposed a distributed solution for network-based model identification, data is shared with an aggregator that performs clustering for model inference. Therefore, no privacy in preserved in this solution.

Moreover, datasets available in the literature for device type or model identification are focused in dimensions such as network connection \cite{aksoy2019automated} or radio frequency fingerprinting \cite{abbas2021improving}. However, there are not execution time-based datasets modeling device performance for identification, just some benchmark datasets focused in other tasks \cite{varghese2021benchmarking}.

Concerning adversarial ML in FL, the authors of \cite{li2020learning} exposed the impossibility of the central server to control the clients of the federated network. A malicious client could send poisoned model updates to the server in order to worsen learning performance. A new framework for federated learning is proposed in which the central server learns to detect and remove malicious model updates using a detection model. Finally, the authors of \cite{rey2021federated} considered the presence of adversaries in their solution for FL-based network attack detection. However, no model identification experiments were carried out.

In conclusion, although each research topic, namely hardware based model identification, federated learning, and adversarial ML, has been separately explored. To the best of our knowledge, and as \tablename~\ref{tab:related} shows, there is no work in the literature analyzing device model identification from a federated learning perspective. Besides, there is not a dataset focused on model identification based on execution time-based features. Furthermore, there is no solution evaluating the impact of adversarial attacks when some clients are malicious, together with the main aggregation-based attack mitigation techniques.

\begin{table*}[htpb!]
    \centering
    \footnotesize
    \caption{Comparison of the most relevant model identification literature works.}
    \begin{tabular}{ C{1cm} C{3cm} C{1.8cm}  C{2cm} C{6cm} } 
    \hline
    \textbf{Work} & \textbf{Model Identification} & \textbf{FL} & \textbf{Adv. attack} & \textbf{Conclusions} \\
    \hline
    \cite{babun2021cps} & \cmark (Hardware-based) & \xmark & \xmark & 0.9873 average accuracy using correlation-based algorithms to recognize 11 device classes. \\
    \hline
    \cite{sanchez2018clock} & \cmark (Hardware-based) & \xmark & \xmark & +200 computers individually identified based on execution-time statistical comparison. \\
    \hline
    \cite{he2021edge} & \cmark (Network-based) & \cmark & \xmark & 0.882 accuracy using a federated LSTM network to identify 10 IoT device types. \\
    \hline
    \cite{mun2021internet} & \textbf{--} (App identification) & \cmark & \xmark & 0.92 accuracy using a federated CNN to identify user-level applications.\\
    \hline
    \cite{Thangavelu2018DEFT} & \cmark  (Network-based) & \xmark (Distributed) & \xmark & $\approx$0.97 accuracy for clustering-based IoT device type classification. \\
    \hline
    This work  &  \cmark (Hardware-based) & \cmark & \cmark(Label-flipping) & 0.9999 accuracy identifying RPi models and adversarial impact analysis. \\
    \hline
    \end{tabular}
    \label{tab:related}
\end{table*}

\section{Scenario and Dataset Creation}
\label{sec:dataset}

This section describes the scenario and the procedure followed to generate the execution time dataset used in the present work. Besides, it provides some insights on the data distribution that can be useful to understand the model identification performance.

\subsection{Scenario description}

In total, a setup of 55 Raspberry Pis from different models but identical software images are employed for data collection, running using \textit{Raspbian 10 (buster) 32 bits} as OS and \textit{Linux kernel 5.4.83}. The generated dataset is composed of 2.750.000 vectors (55 devices * 50000 vectors per device). Each vector has two labels associated, one regaring the individual device that generated it, and another regarding the model of this device. Data collection was performed under normal device functioning and default frequency and power configuration, where the CPU frequency is automatically adjusted according to the workload. The list of devices contained in the dataset is showed in \tablename~\ref{tab:device}.

\begin{table}[pos=htpb!]
    \centering
    \caption{Devices employed in data collection.}
    \begin{tabular}{ P{1.8cm} P{2.7cm} P{1.7cm} } 
    \hline
    \textbf{N devices} & \textbf{Model} & \textbf{N samples} \\
    \hline
    12 & RPi 4 Model B & 660000 \\
    22 & RPi 3 Model B+ & 1210000 \\
    5 & RPi 2 Model B & 275000 \\
    16 & RPi 1 Model B & 880000 \\
    \hline
    \end{tabular}
    \label{tab:device}
\end{table}

\subsection{Dataset creation}

The generated dataset has been made publicly available \cite{bovet2021dataset} for download and research of other authors. The published data includes both identifiers for RPi model and for individual devices, so new research could be done regarding individual device identification.

For the device performance dataset generation, the CPU performance of the device was leveraged as data source. In this sense, the time to execute a software-based random number generation function was measured in microseconds.

To minimize the impact of noise and other processes running in the device, the monitored function was executed in groups of 1000 runs a total number of 50000 times per group. Then, for each 1000-run group, a set of statistical features was calculated, generating a performance fingerprint composed of 50000 vectors per device. In total, 13 statistical features are calculated: maximum, minimum, mean, median, standard deviation, mode sum, minimum decrease, maximum decrease, decrease summation, minimum increase, maximum increase and increase summation. Decrease and increase values are calculated as the negative or positive difference between two consecutive values in each 1000-run group. Besides, the device model is added as label. \tablename~\ref{tab:vector} shows an example of a vector in the dataset belonging to a Raspberry Pi 4 device.

\begin{table*}[pos=htpb!]
    \centering
    \footnotesize
    \caption{Dataset vector example for a RPi4 device. Values represent the time required to execute a function expressed in microseconds.}
    \begin{tabular}{ P{0.5cm} P{0.5cm}  P{0.7cm}  P{0.95cm}  P{0.6cm}  P{0.7cm}  P{0.8cm}  P{1cm}  P{0.8cm}  P{0.8cm}  P{0.6cm}  P{0.6cm} P{0.6cm} P{0.8cm} } 
    \hline
    \textbf{Min} & \textbf{Max} & \textbf{Mean}& \textbf{Median}& \textbf{Std Dev}& \textbf{Mode}& \textbf{Sum}& \textbf{Min Decr.}& \textbf{Max Decr.}& \textbf{Decr. sum}& \textbf{Min Incr.}& \textbf{Max Incr.} & \textbf{Incr. sum} & \textbf{Model} \\
    \hline
    2.1 & 12.7 & 4.2 & 4.5 & 1.3 & 3.5 & 4221.8 & -113.74 & -8.7 & -0.001 & 117.8 & 0.001 & 7.81 & RPi4\\
    \hline
    \end{tabular}
    \label{tab:vector}
\end{table*}

\subsection{Data exploration}

\begin{figure}[htpb!]
    \centering
    \includegraphics[width=\columnwidth]{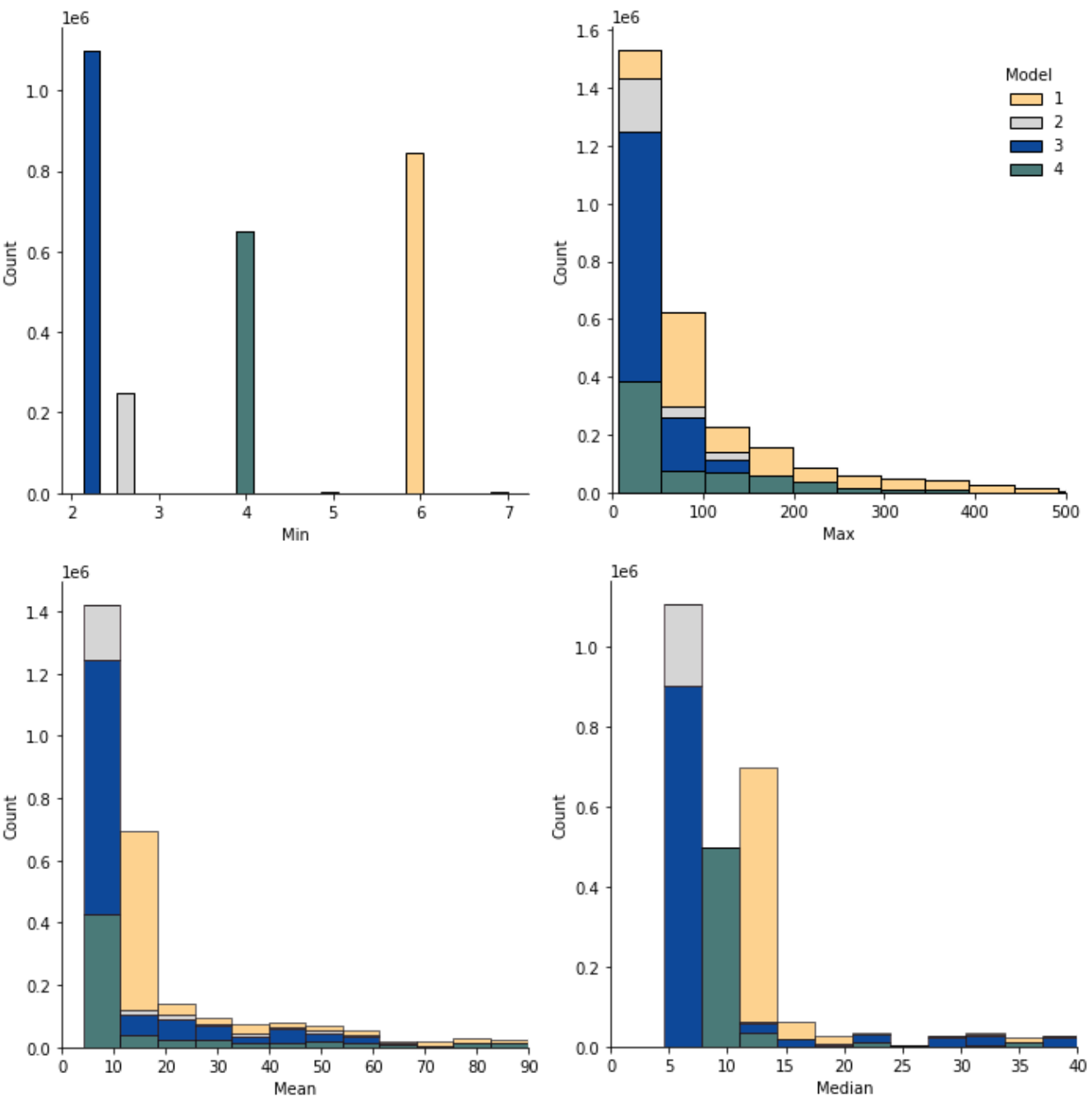}
    \caption{Min, Max, Mean and Median feature distributions}
    \label{fig:distribution}
\end{figure}

\figurename~\ref{fig:distribution} shows the data distribution for min, max, mean and median features. It can be observed how the values vary according to the model that generated the vector, resulting in a presumably good model identification performance.

\section{Centralized vs Federated Model Identification Performance}
\label{sec:classification}

This sections seeks to evaluate firstly the performance of the generated dataset when identifying the different device models in a DL-based centralized setup, and secondly the performance variation when the model is generated in a distributed manner, following a FL-based approach.

\subsection{Centralized setup}

For the centralized experiment, the dataset described in Section \ref{sec:dataset} is divided in 80\% for training/validation and 20\% for testing, without data suffling. \textit{Min-max normalization} is applied then using the training data to set the boundaries.

\begin{equation}
    x'=\frac{x-min(x)}{max(x)-min(x)}
\end{equation}

To measure the centralized classification performance, a (\textit{MLP}) classifier is implemented. After several iterations testing different number of layers and neurons per layer, the chosen MLP architecture is composed of 13 neurons in the input layer (one per feature), two hidden layers with 100 neurons each one using \textit{relu (Rectified Linear Unit)} as activation function \cite{agarap2018deep}, and 4 neurons in the output \textit{softmax} layer (one per model class). \textit{Adam} \cite{kingma2014adam} was used as optimizer with a 0.001 learning rate, and 0.9 and 0.999 as first and second-order moments. \tablename~\ref{tab:model} shows the details of the model.

\begin{table}[pos=htpb!]
    \centering
    \caption{MLP architecture for model identification.}
    \begin{tabular}{ P{1.5cm} P{2cm} P{2.7cm} } 
    \hline
    \textbf{Layer} & \textbf{Neurons} & \textbf{Activation} \\
    \hline
    1 & 13 & - \\
    2 & 100 & relu \\
    3 & 100 & relu \\
    4 & 4 & softmax \\
    \hline
    \textbf{Optimizer} & Adam \\
    \hline
    \end{tabular}
    \label{tab:model}
\end{table}

With this setup, the MLP is trained for 100 epochs using \textit{early stopping} if no validation accuracy improvement occurs in 20 epochs.

\begin{figure}[htpb!]
    \centering
    \includegraphics[width=0.9\columnwidth]{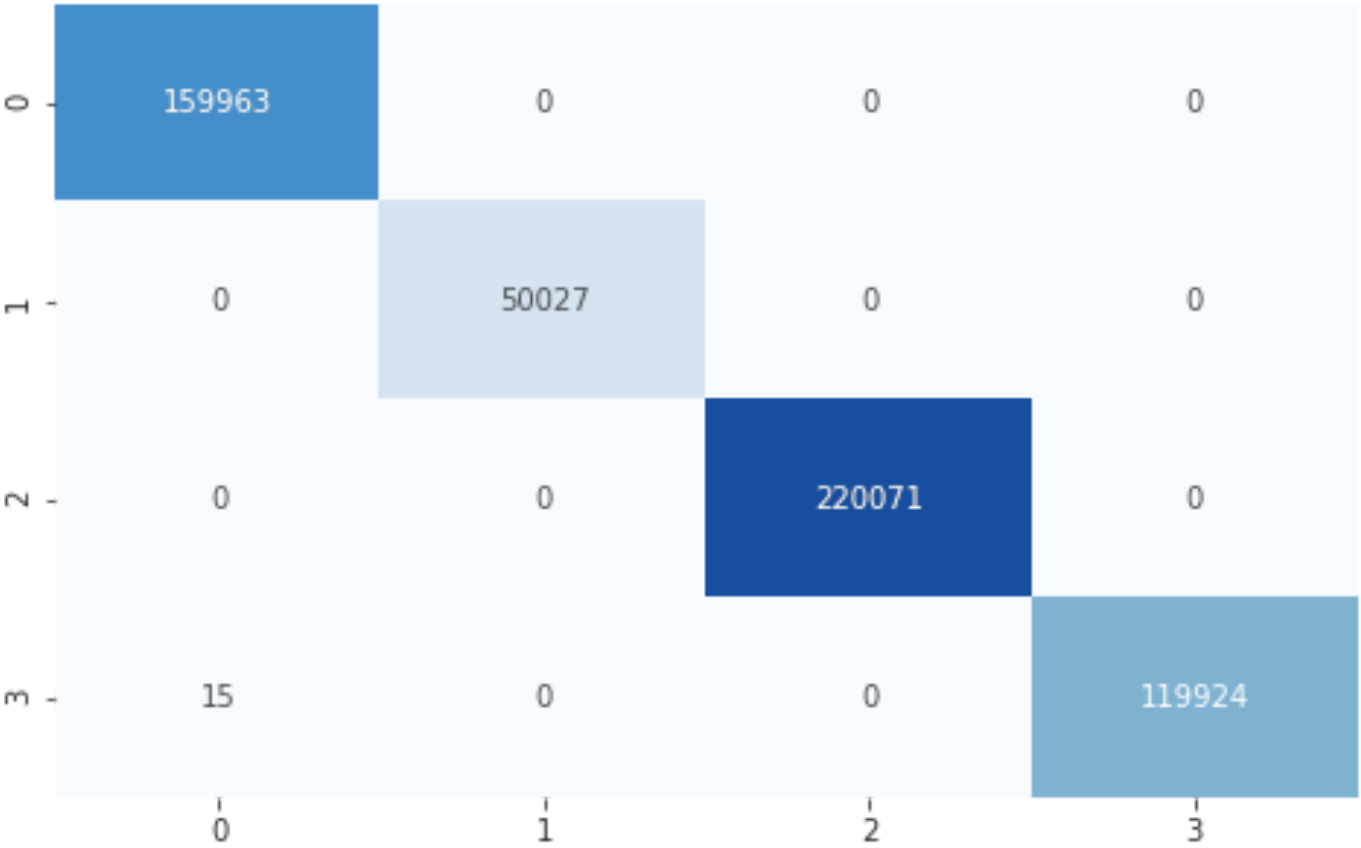}
    \caption{Centralized evaluation confusion matrix.}
    \label{fig:centralized}
\end{figure}

\figurename~\ref{fig:centralized} shows the confusion matrix resultant of the evaluation of the test datastet. As it can be seen, almost a perfect identification is achieved, with only 15 samples being misclassified out of $\approx$550000 (0.999972 accuracy). These results are aligned with the expectations, as having different CPUs in each RPi model makes the execution time of the same functions different between them. However, model identification performance is not the main focus of the present work, where the priority is to prove the effectiveness of a federated setup and the impact of adversarial attacks and countermeasures.

\subsection{Federated scenario and results}

Once the centralized model has been obtained, the decentralized model is implemented using FL to compare the performance of both approaches. The FL approach is based in horizontal FL, where the clients have datasets with the same features but from different data samples.

For implementation, the IBM Federated Learning library \cite{ibmfl2020ibm} is used, which incorporates the necessary tools to perform the training in a decentralized manner.

\subsubsection{Scenario}

\begin{figure}[htpb!]
    \centering
    \includegraphics[width=\columnwidth]{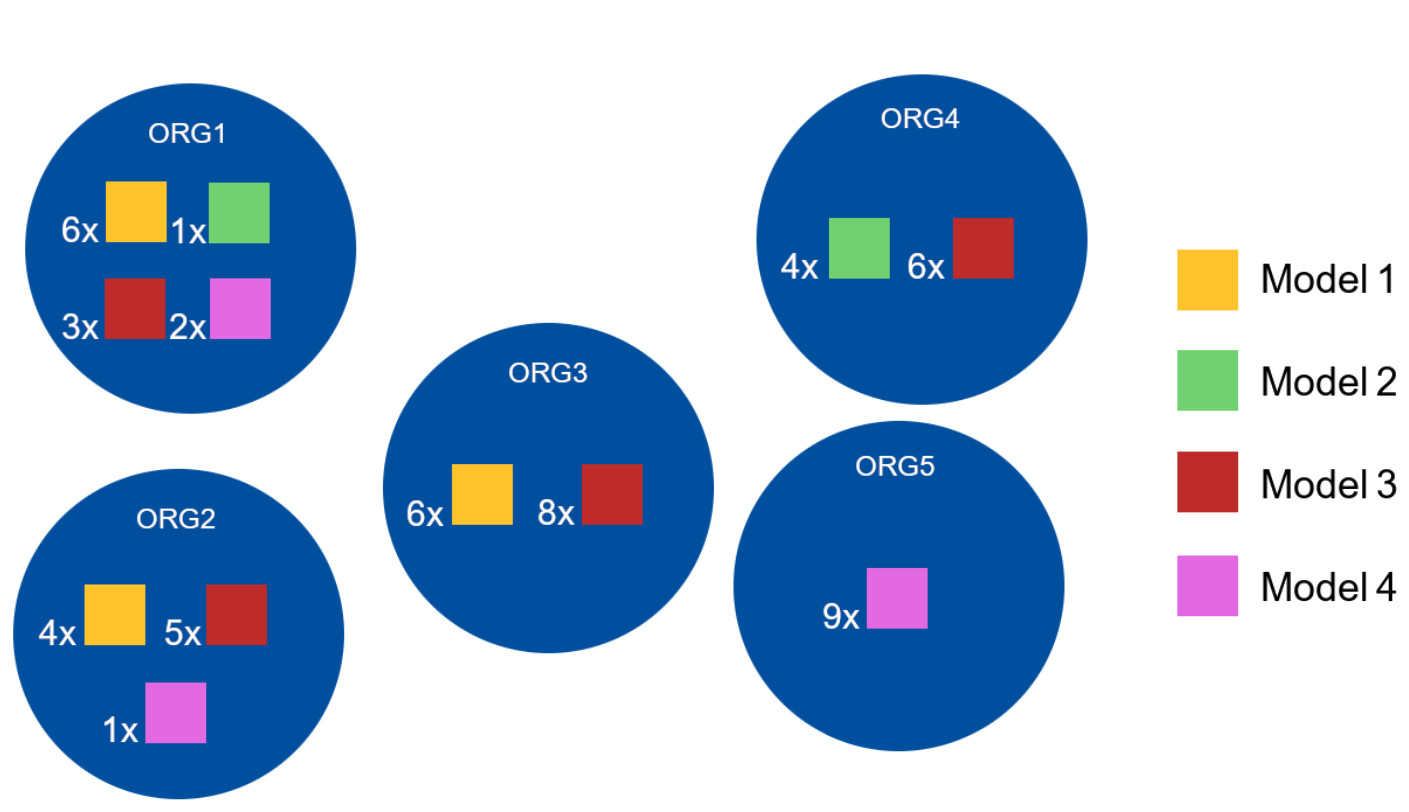}
    \caption{Data division in the federated learning scenario.}
    \label{fig:fedscenario}
\end{figure}

For the decentralization of the training phase, a scenario has been created in which there are 5 independent organizations in which the available data are distributed. Each of them has a certain number of devices belonging to different models, but not all of them have information on all models, i.e. there are organizations that only have devices of type 4 model, others that only have devices of type 2 and 3 models, etc. \figurename~\ref{fig:fedscenario} provides the details of the device distribution in each organization. Therefore, the 5 organizations intend to generate a global model capable of identifying all the existing device models among all of them. This setup leads to an scenario of Non-IID (Non-Independent and Identically Distributed) data, harder to solve with FL as model aggregation will be negatively influenced in the aggregated models are very different to each other.

\subsubsection{Federated architecture design}

\begin{figure}[htpb!]
    \centering
    \includegraphics[width=\columnwidth]{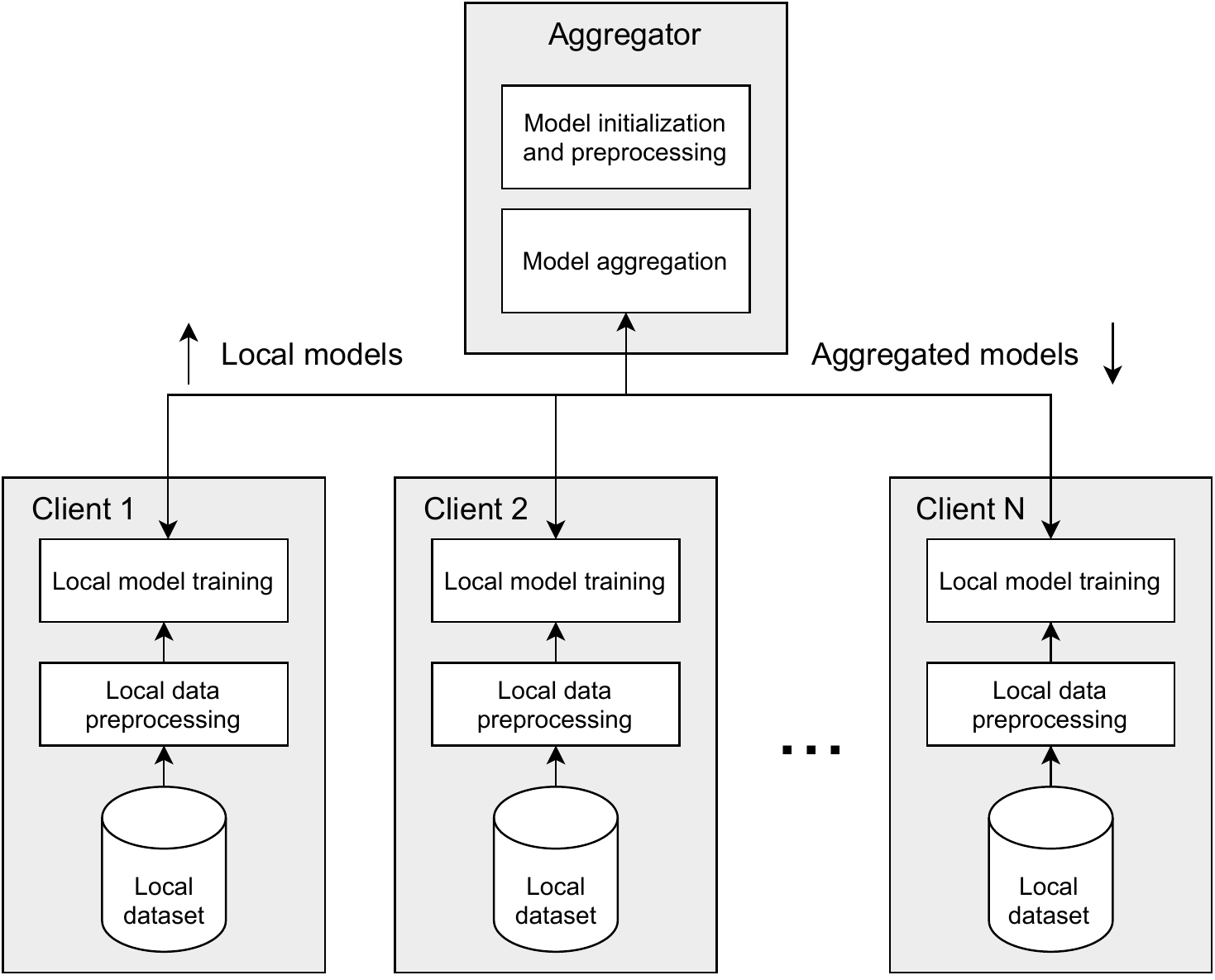}
    \caption{Designed federated architecture for experimentation.}
    \label{fig:architecture}
\end{figure}

In order to test the performance of a FL-based setup, first it is necessary to define the architecture to be implemented. In this sense, \figurename~\ref{fig:architecture} shows the organization of the different client which will hold the data and upload their local models to the aggregator in order to cyclically build a common model capable of making predictions based on the local data of all clients.

\subsubsection{Performance Evaluation}

In order to fairly compare the models, the MLP architecture to be trained will be the same as the one used in the centralized model (see \tablename~\ref{tab:model}, i.e. the layers will have 13, 100, 100, 100, and 4 neurons, from input to output. As aggregation method, Federated Averaging is applied as proposed in \cite{mcmahan2017communicationefficient}.  As initialization step, the aggregation server performs two tasks: (1) to initialize the weights of the model that the clients will start to train, so all clients start from the same setup; (2) to retrieve for each client its min-max values of each feature for common dataset normalization, having a \textit{min-max normalization} for each dataset \textit{x} in organization \textit{o} $\in x$ defined as:

\begin{equation}
    x'_o=\frac{x_o-min(x_{i\in[n]})}
    {max(x_{i\in[n]})-min(x_{i\in[n]})}
\end{equation}

Algorithm \ref{alg:fedavg} defines the iterative training process for the model generation, assuming previous dataset normalization. Each client performs local updates of the model and returns them to the server for aggregation, repeating then the process for the desired number of rounds.

\begin{algorithm}[ht!]
\begin{algorithmic}
 \item \textbf{Server executes:}
 \item  \hskip1em initialize $w_0$ 
 
 \item  \hskip1em \textbf{for} each round \textit{t} = 1,2... \textbf{do}
 \item  \hskip2em $m\leftarrow max(C \cdot K,1)$
 \item  \hskip2em $S_t \leftarrow$ (random set of $m$ clients)
 \item  \hskip2em \textbf{for} each client $k \in S_t$ \textbf{in parallel do}
 
 \item  \hskip3em $w^k_{t+1} \leftarrow $ ClientUpdate($k,w_t$)
 
 \item  \hskip2em $w_{t+1} \leftarrow \sum_{k=1}^{K} \frac{n_k}{n} w_{t+1}^k$ //Aggregation\\

 \item  \textbf{ClientUpdate(}$k,w$\textbf{):} //\textit{Run on client} $k$ 

 \item  \hskip1em $B \leftarrow$ (split $P_k$ into batches of size $B$) 
 
 \item  \hskip1em \textbf{for} each local epoch \textit{i} from 1 to \textit{W} \textbf{do}
 \item  \hskip2em \textbf{for} batch \textit{b} $\in B$ \textbf{do}

  \item \hskip3em $w \leftarrow w - \eta \bigtriangledown l(w;b)$ //Local update 

 \item \hskip1em return $w$ to server

 \end{algorithmic}
 \caption{\texttt{FederatedAveraging}. The \textit{K} clients are indexed by \textit{k}; \textit{B} is the local minibatch size, \textit{E} is the number of local epochs, and $\eta$ is the learning rate; $w$ are the model weights; $P_k$ is the local dataset of client $k$. \cite{mcmahan2017communicationefficient}}
 \label{alg:fedavg}
\end{algorithm}

\begin{figure}[htpb!]
    \centering
    \includegraphics[width=\columnwidth]{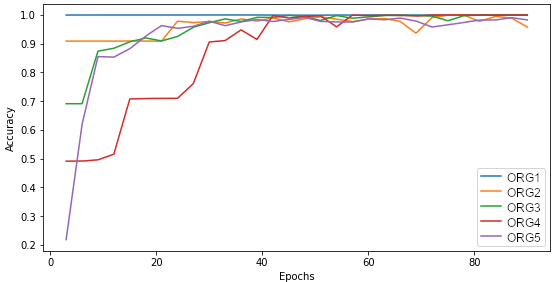}
    \caption{Federated training validation accuracy evolution.}
    \label{fig:fedtraining}
\end{figure}

The training process was executed for 90 federated rounds, with one epoch per round. \figurename~\ref{fig:fedtraining} shows the evolution of the local validation accuracy for each one of the clients during the training process. It can be appreciated how the maximum performance is reached around epoch 50, and then the accuracy scores for each client keep oscillating between 0.95 and 1 until round 90.

\begin{figure}[htpb!]
    \centering
    \includegraphics[width=0.9\columnwidth]{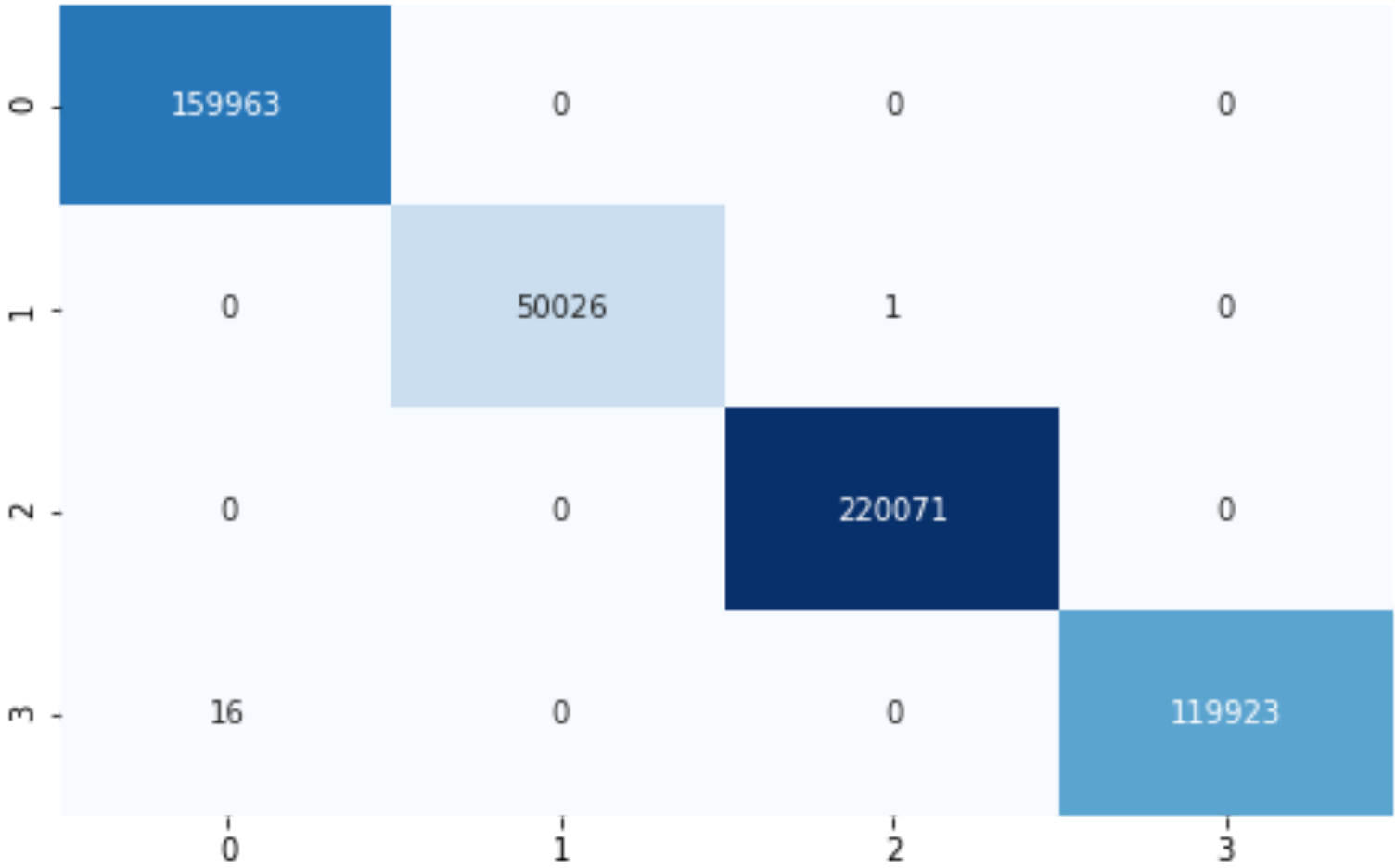}
    \caption{Federated evaluation confusion matrix.}
    \label{fig:federated}
\end{figure}

Regarding performance, \figurename~\ref{fig:federated} shows the results of the test dataset evaluation, the same dataset than in the centralized setup. Here, the results are almost identical, with only 17 errors in $\approx$550000 test samples and an accuracy of 0.999969.

From the previous results, a main conclusion can be extracted: no performance loss has been introduced in the resultant model due to the application of a FL-based approach. Besides, as no data has left each organization in the process, the privacy of the information has been kept private successfully.

\section{Adversarial Attack and Robust Aggregation}
\label{sec:adversarial}

After testing the effectiveness of Federated Learning, its robustness will be tested using adversarial attacks, specifically the label-flipping technique, using different aggregation algorithms in order to see which one best fits the proposed scenario in the presence of attacks.

\subsection{Label flipping attack}

The label-flipping adversarial technique is applied during the training process, using the same scenario described above with the difference that this time part of the data will be poisoned.

In this sense, the federated training is carried out by poisoning 25, 50, 75 and 100\% of the data of 1, 2 and 3 different organizations, representing 20\%, 40\% and 60\% malicious clients, respectively. These configurations are used because potential malicious clients may not poison all their data and just one portion, in order to go undetected and make their activity more difficult to identify. So, a total of 12 adversarial scenarios have been created (4 poising percentages * 3 possible malicious organizations). This setup is generated by modifying the labels of the training data, changing the value of each label to a random value between 1 and 4 that is not the value of the original label. The poisoned organizations are ORG1, ORG2 and ORG4 (in that order for 1, 2, 3 malicious clients).

\begin{figure}[htpb!]
    \centering
    \includegraphics[width=\columnwidth]{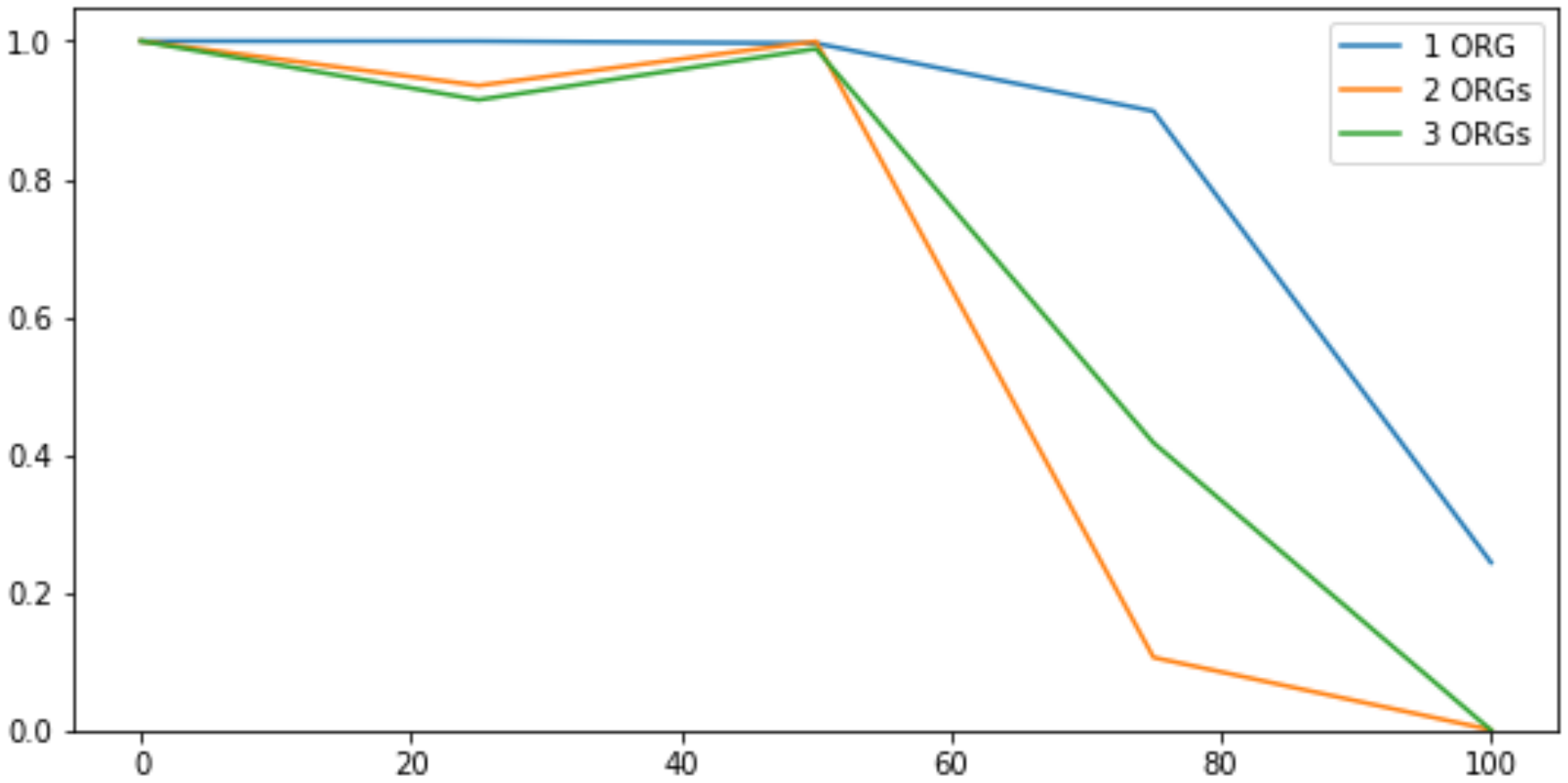}
    \caption{FedAvg performance against label-flipping attack. (X axis depicts poisoning percentage, Y axis depicts accuracy).}
    \label{fig:fedavglabel}
\end{figure}

\figurename~\ref{fig:fedavglabel} shows the results when FedAvg is applied as aggregation algorithm in the 12 previous adversarial scenarios (as well as when no label-flipping attack is applied).

As can be seen, aggregation by averaging offers good performance up to a 50\% poisoning, maintaining the accuracy over 0.9. However, accuracy drops rapidly to hit rates close to 0\% when the poisoning is 75\% or higher. Therefore, FedAvg cannot be considered a robust aggregation method in the presence of the label-flipping attack. Next, 3 different aggregation methods will be analyzed in the following in order to check which one offers better performance.

\subsection{Robust aggregation methods}

Next, several aggregation methods focused on improving the model resilience to malicious clients will be evaluated and compared to the default FedAvg algorithm.

\subsubsection{Coordinate-wise median aggregation}

Coordinate-wise median \cite{yin2018byzantine} follows the scheme of the aggregation by average with the difference that the combination of the weights is done by calculating the median of each weight of the local models. In short, following Algorithm \ref{alg:fedavg}, the averaging aggregation step is substituted by a median operation.

\begin{figure}[htpb!]
    \centering
    \includegraphics[width=\columnwidth]{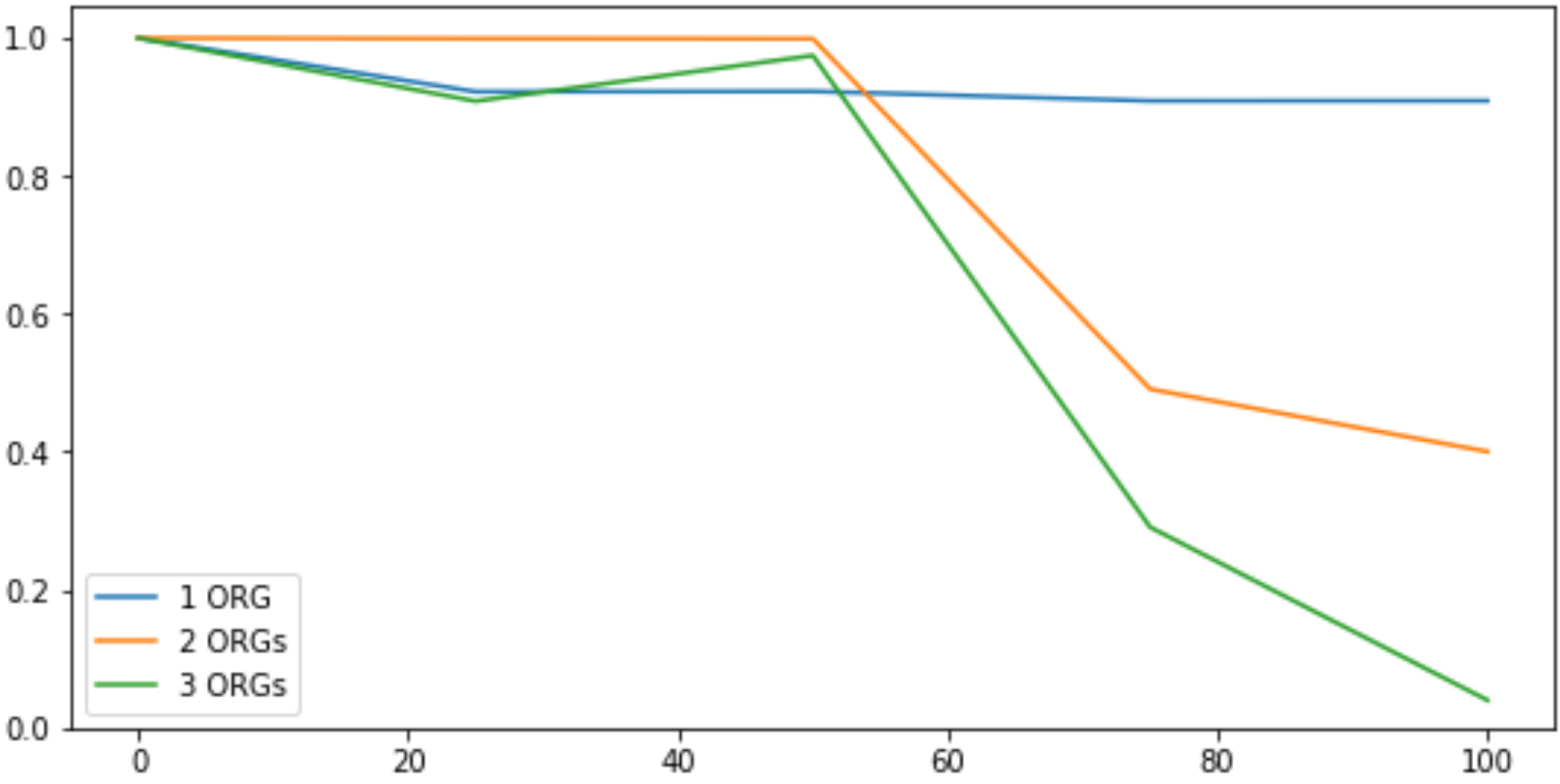}
    \caption{Coordinate-wise median aggregation performance against label-flipping attack. (X axis depicts poisoning percentage, Y axis depicts accuracy).}
    \label{fig:medianlabel}
\end{figure}

Next, \figurename~\ref{fig:medianlabel} depicts the accuracy results when the different attack setups are applied when using median aggregation. Coordinate-wise median follows a similar pattern to FedAvg aggregation, dropping from 50\% poisoning rate. However, it has performed better especially when there is only one poisoned organization (20\% malicious clients). While FedAvg dropped to 0.20-0.40, the median has remained around $\approx$0.9.

\subsubsection{Krum aggregation}

\begin{figure}[htpb!]
    \centering
    \includegraphics[width=\columnwidth]{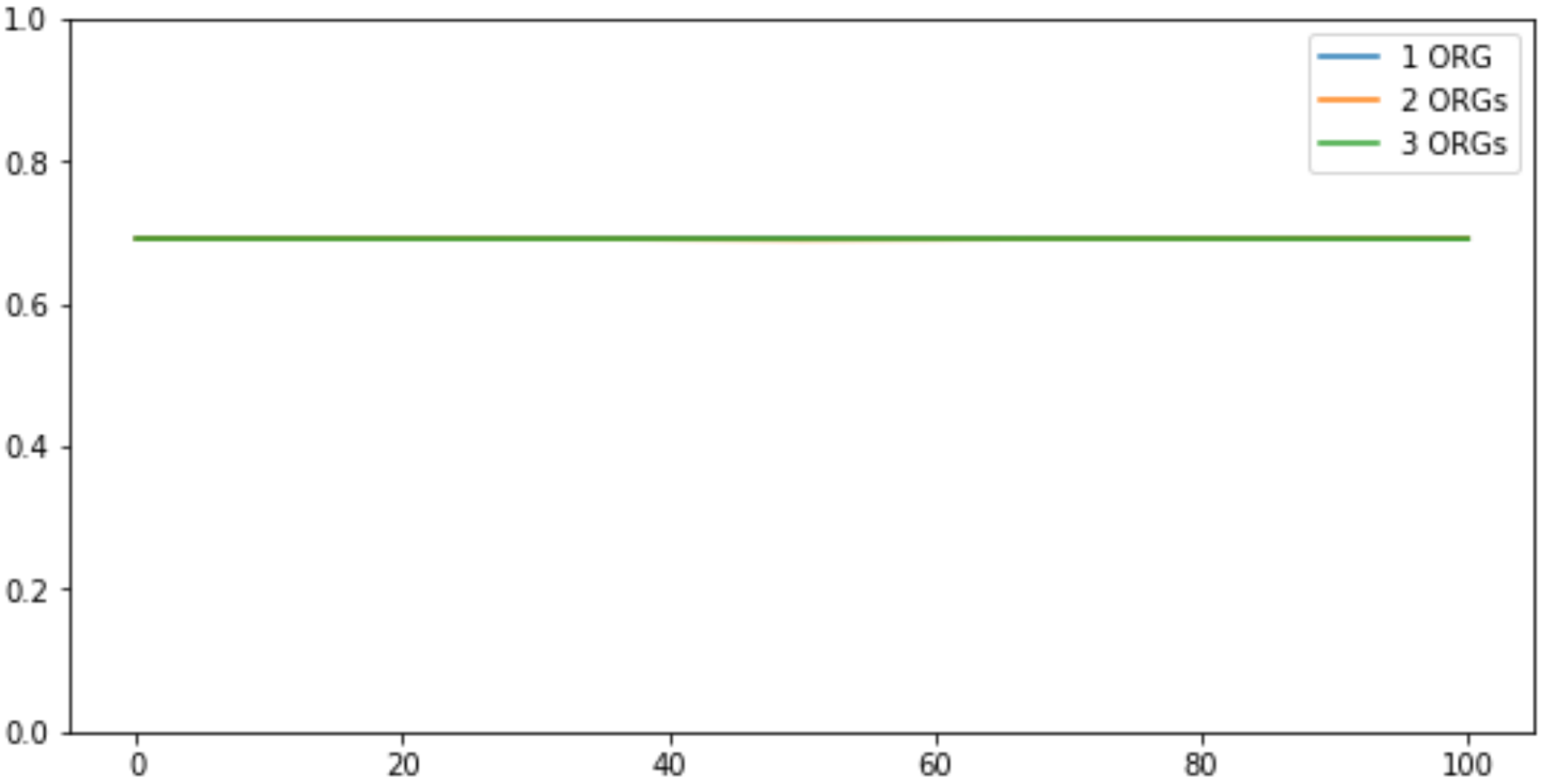}
    \caption{Krum performance against label-flipping attack. (X axis depicts poisoning percentage, Y axis depicts accuracy).}
    \label{fig:krumlabel}
\end{figure}

The idea behind Krum \cite{blanchard2017machine} is to select one of the \textit{m} local models that is most similar to the rest as the global model. The idea is that even if the selected model is a poisoned model the impact would not be so great since it would be similar to other models that are probably not poisoned. The aggregator calculates the sum of the distances between each model and its closest local models. Krum selects the local model with the smallest sum of distances as the global model. \figurename~\ref{fig:krumlabel} shows the results when Krum is applied as aggregation algorithm.

\begin{figure}[htpb!]
    \centering
    \includegraphics[width=0.9\columnwidth]{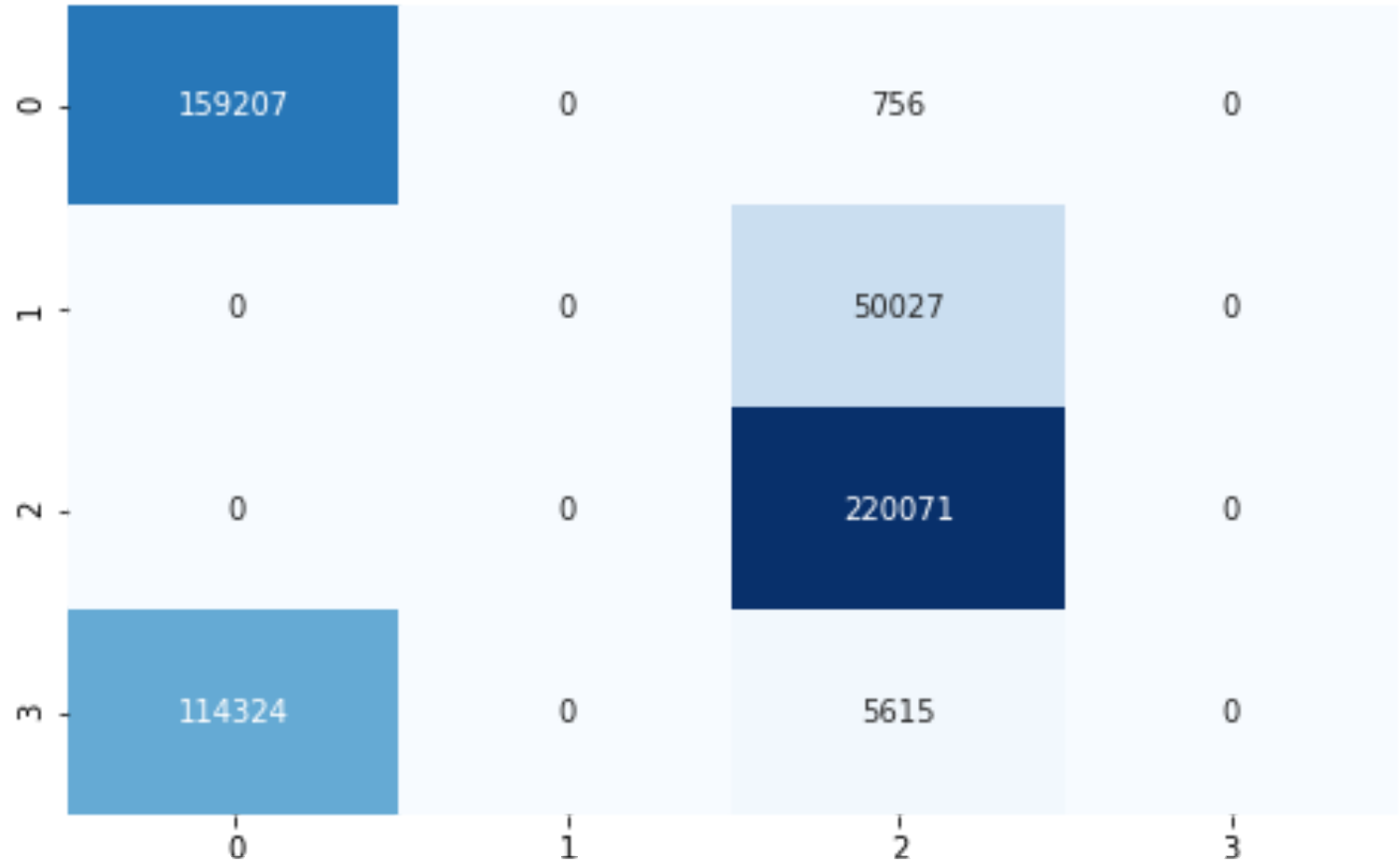}
    \caption{Krum confusion matrix during label-flipping attack.}
    \label{fig:krummatrix}
\end{figure}

As can be seen, Krum has remained constant for all configurations with an accuracy of 0.6896 . This is because this aggregation method chooses a single local model as the global model and discards the information from the rest of the local models. Therefore, what is happening is that it always chooses the same local model, and this one belongs to an organization that has not been poisoned, so the hit rate remains constant. \figurename~\ref{fig:krummatrix} shows that the resulting global model only recognizes device models of types 0 and 2. 

On the other hand, this organization has not been poisoned, which explains that the performance remains constant since the resulting global model is identical regardless of the percentage of poisoning. Therefore, it can be concluded that Krum is selecting the resulting local model of organization 3 in all scenarios, loosing the information regarding the classes not seen in this organization (see \figurename~\ref{fig:fedscenario}).

\subsubsection{Zeno aggregation}

\begin{figure}[htpb!]
    \centering
    \includegraphics[width=\columnwidth]{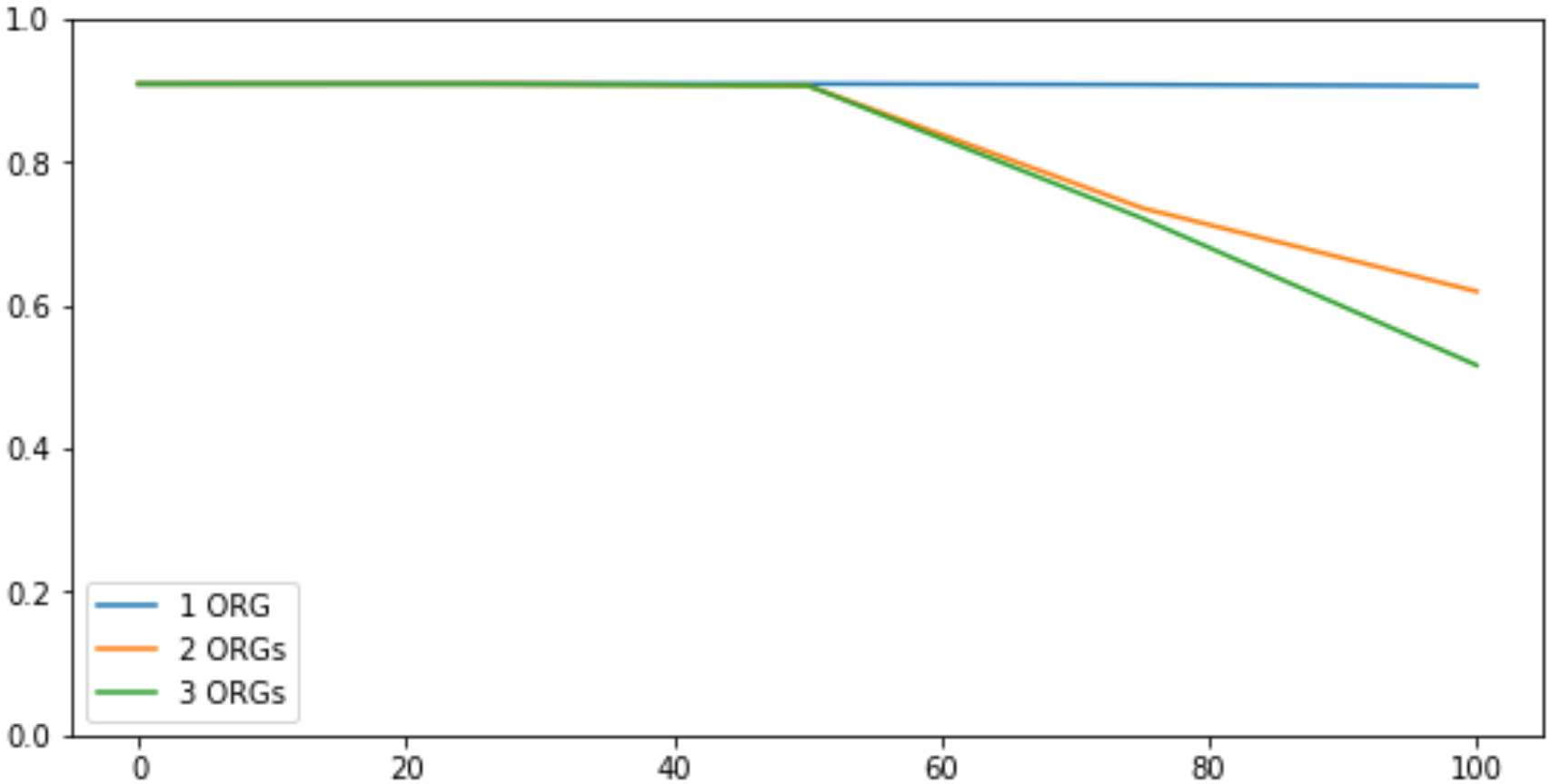}
    \caption{Zeno performance against label-flipping attack. (X axis depicts poisoning percentage, Y axis depicts accuracy).}
    \label{fig:zenolabel}
\end{figure}

Zeno \cite{xie2019zeno} is suspicious of potentially malicious organizations and uses a ranking-based preference mechanism. The number of malicious organizations can be arbitrarily large, and only the assumption that 'clean' organizations exist (at least one) is used. Each organization is ranked based on the estimated descent of the loss function. The algorithm then aggregates the organizations with the highest scores. The score roughly indicates the reliability of each organization. In this sense, it could be seen as a combination of Krum and averaging aggregation mechanisms. \figurename~\ref{fig:zenolabel} shows the results when Zeno is applied as aggregation algorithm.

\begin{figure}[htpb!]
    \centering
    \includegraphics[width=0.9\columnwidth]{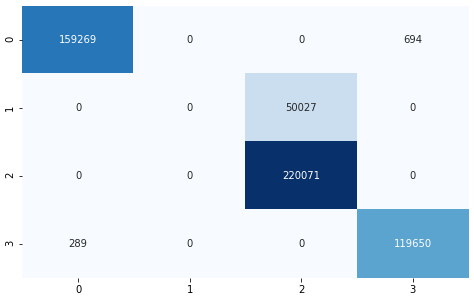}
    \caption{Zeno confusion matrix with one malicious client.}
    \label{fig:zenomatrix}
\end{figure}

In this case, Zeno has outperformed the aggregation by median and Krum when only one client is malicious (20\% of the total), achieving 0.9072 accuracy. When there is only one poisoned organization Zeno remains constant without being altered by this attack. \figurename~\ref{fig:zenomatrix} shows the confusion matrix of Zeno when one client is malicious. It can be appreciated how the performance decrease comes from the impossibility of classifying the second class, the one under represented in the scenario as there are only 5 RPi2 in the dataset.

\begin{figure}[htpb!]
    \centering
    \begin{subfigure}{0.50\textwidth}
    \begin{center}
    \includegraphics[width=\columnwidth]{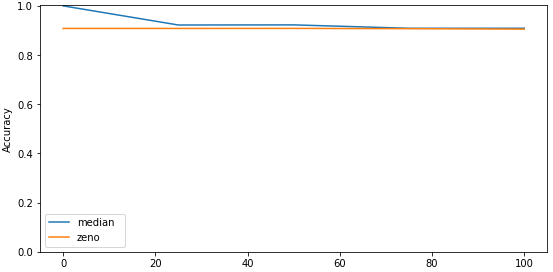}
    \end{center}
    \caption{One organization poisoned.}
    \end{subfigure}
    \begin{subfigure}{0.49\textwidth}
    \begin{center}
    \includegraphics[width=\columnwidth]{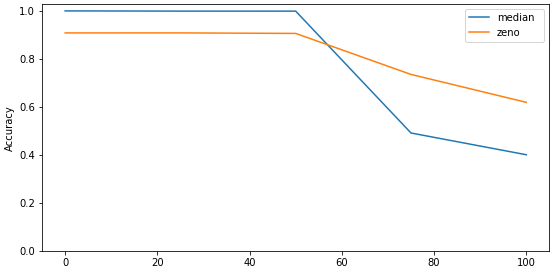}
    \end{center}
    \caption{Two organizations poisoned.}
    \end{subfigure}
    \begin{subfigure}{0.49\textwidth}
    \begin{center}
    \includegraphics[width=\columnwidth]{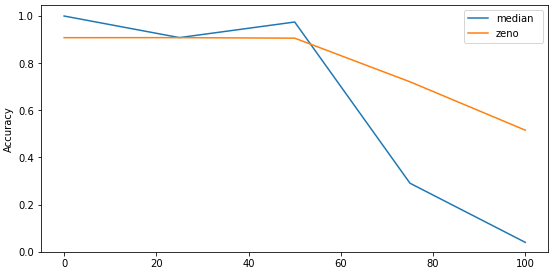}
    \end{center}
    \caption{Three organizations poisoned.}
    \end{subfigure}
    \caption{Zeno and coordinate-wise median aggregation with different number of poisoned clients (X axis depicts poisoning percentage, Y axis depicts accuracy).}
    \label{fig:zenomed}
\end{figure}

When there are 2 or 3 poisoned organizations, Zeno performance drops once the poisoning rate reaches 75\% and 100\%, but it still manages to maintain an acceptable performance above 0.50, considering the degree of the attack. \figurename~\ref{fig:zenomed} compares the performance evolution of Zeno and coordinate-wise median with different number of poisoned organizations. As it can be appreciated, median performance is higher in all scenarios until the poisoning percentage goes above 50\%. After that, Zeno shows a better or equal performance in all cases, being the greater difference when three organizations are completely malicious (60\% malicious clients).

\section{Conclusions and Future Work}
\label{sec:conclusions}

In the present work it has been demonstrated that it is possible to identify device models using only statistical data concerning the CPU execution time of the device. An MLP model has been obtained capable of identifying four RPi device models with a 99.99\% accuracy rate.
Besides, the effectiveness of Federated Learning technique has been tested against centralized learning. For this setup, a scenario has been proposed where a total of 5 organizations aim to create a model capable of identifying the device models without sharing the actual data with each other. The resulting model has obtained identical performance in both cases, centralized and distributed. Thus taking advantage of the benefits offered by Federated Learning, training a data privacy and data security preserving model, while maintaining the efficiency of the model obtained through a traditional approach.
On the other hand, different aggregation algorithms have been tested in order to check which one best fits the proposed scenario facing a label-flipping attack. Zeno has turned out to be the best performing aggregation method in the presence of attacks due to combining the Krum and mean aggregation methods. By selecting the m best models and aggregating them using mean aggregation, less information is lost than with Krum by ignoring certain organizations that are considered malicious. Finally, the data collected for the previous experimentation has been made publicly available due to the lack of performance fingerprinting datasets focused on device identification, and prepared for FL-based setups.

As future work, the efforts will be focused on experimentation with more types of device models with more complex scenarios such as making each device a single client instead of being grouped into organizations. On device identification, it is planned to focus on identifying individual devices with a high hit rate and not just identifying device models, as well as testing other modes of identification by collecting data from other hardware elements than the CPU. It would also be interesting to poison the local model weights instead of the local data (model poisoning) or experiment with other adversarial attack techniques such as Evasion attacks, where the goal is to trick the model once it is trained and not to poison the training process.

\section*{Data Availability}

The datasets generated during and/or analysed during the current study are available in the RPi model device identification Mendeley Data repository, \\ https://doi.org/10.17632/vr9wztmfxg.2

\section*{Acknowledgment}

This work has been partially supported by \textit{(a)} the Swiss Federal Office for Defense Procurement (armasuisse) with the TREASURE (R-3210/047-31) and CyberSpec (CYD-C-2020003)  projects and \textit{(b)} the University of Zürich UZH.

\bibliographystyle{cas-model2-names}
\bibliography{sn-bibliography}

\begin{thebibliography}{25}
\expandafter\ifx\csname natexlab\endcsname\relax\def\natexlab#1{#1}\fi
\providecommand{\url}[1]{\texttt{#1}}
\providecommand{\href}[2]{#2}
\providecommand{\path}[1]{#1}
\providecommand{\DOIprefix}{doi:}
\providecommand{\ArXivprefix}{arXiv:}
\providecommand{\URLprefix}{URL: }
\providecommand{\Pubmedprefix}{pmid:}
\providecommand{\doi}[1]{\href{http://dx.doi.org/#1}{\path{#1}}}
\providecommand{\Pubmed}[1]{\href{pmid:#1}{\path{#1}}}
\providecommand{\bibinfo}[2]{#2}
\ifx\xfnm\relax \def\xfnm[#1]{\unskip,\space#1}\fi
\bibitem[{Abbas et~al.(2021)Abbas, Nasir, Nouichi, Abdelsalam, Talib, Waraga
  et~al.}]{abbas2021improving}
\bibinfo{author}{Abbas, S.}, \bibinfo{author}{Nasir, Q.},
  \bibinfo{author}{Nouichi, D.}, \bibinfo{author}{Abdelsalam, M.},
  \bibinfo{author}{Talib, M.A.}, \bibinfo{author}{Waraga, O.A.}, et~al.,
  \bibinfo{year}{2021}.
\newblock \bibinfo{title}{Improving security of the internet of things via rf
  fingerprinting based device identification system}.
\newblock \bibinfo{journal}{Neural Computing and Applications} ,
  \bibinfo{pages}{1--17}.
\bibitem[{Agarap(2018)}]{agarap2018deep}
\bibinfo{author}{Agarap, A.F.}, \bibinfo{year}{2018}.
\newblock \bibinfo{title}{Deep learning using rectified linear units (relu)}.
\newblock \bibinfo{journal}{arXiv preprint arXiv:1803.08375} .
\bibitem[{Aksoy and Gunes(2019)}]{aksoy2019automated}
\bibinfo{author}{Aksoy, A.}, \bibinfo{author}{Gunes, M.H.},
  \bibinfo{year}{2019}.
\newblock \bibinfo{title}{Automated iot device identification using network
  traffic}, in: \bibinfo{booktitle}{ICC 2019-2019 IEEE International Conference
  on Communications (ICC)}, \bibinfo{organization}{IEEE}. pp.
  \bibinfo{pages}{1--7}.
\bibitem[{Babun et~al.(2021)Babun, Aksu and Uluagac}]{babun2021cps}
\bibinfo{author}{Babun, L.}, \bibinfo{author}{Aksu, H.},
  \bibinfo{author}{Uluagac, A.S.}, \bibinfo{year}{2021}.
\newblock \bibinfo{title}{Cps device-class identification via behavioral
  fingerprinting: From theory to practice}.
\newblock \bibinfo{journal}{IEEE Transactions on Information Forensics and
  Security} \bibinfo{volume}{16}, \bibinfo{pages}{2413--2428}.
\bibitem[{Blanchard et~al.(2017)Blanchard, El~Mhamdi, Guerraoui and
  Stainer}]{blanchard2017machine}
\bibinfo{author}{Blanchard, P.}, \bibinfo{author}{El~Mhamdi, E.M.},
  \bibinfo{author}{Guerraoui, R.}, \bibinfo{author}{Stainer, J.},
  \bibinfo{year}{2017}.
\newblock \bibinfo{title}{Machine learning with adversaries: Byzantine tolerant
  gradient descent}, in: \bibinfo{booktitle}{Proceedings of the 31st
  International Conference on Neural Information Processing Systems}, pp.
  \bibinfo{pages}{118--128}.
\bibitem[{Bovet and Sánchez~Sánchez(2021)}]{bovet2021dataset}
\bibinfo{author}{Bovet, G.}, \bibinfo{author}{Sánchez~Sánchez, P.M.},
  \bibinfo{year}{2021}.
\newblock \bibinfo{title}{Rpi model device identification}.
\newblock \URLprefix \url{http://dx.doi.org/10.17632/vr9wztmfxg.2},
  \DOIprefix\doi{10.17632/vr9wztmfxg.2}.
\bibitem[{Cisco(2020)}]{cisco_report}
\bibinfo{author}{Cisco}, \bibinfo{year}{2020}.
\newblock \bibinfo{title}{Cisco annual internet report (2018–2023) white
  paper}.
\newblock
  \bibinfo{howpublished}{\url{https://www.cisco.com/c/en/us/solutions/collateral/executive-perspectives/annual-internet-report/white-paper-c11-741490.html}}.
\newblock \bibinfo{note}{[Online; accessed 22-November-2021]}.
\bibitem[{He et~al.(2021)He, Yin, Wang, Gui, Adebisi, Ohtsuki, Gacanin and
  Sari}]{he2021edge}
\bibinfo{author}{He, Z.}, \bibinfo{author}{Yin, J.}, \bibinfo{author}{Wang,
  Y.}, \bibinfo{author}{Gui, G.}, \bibinfo{author}{Adebisi, B.},
  \bibinfo{author}{Ohtsuki, T.}, \bibinfo{author}{Gacanin, H.},
  \bibinfo{author}{Sari, H.}, \bibinfo{year}{2021}.
\newblock \bibinfo{title}{Edge device identification based on federated
  learning and network traffic feature engineering}.
\newblock \bibinfo{journal}{IEEE Transactions on Cognitive Communications and
  Networking} .
\bibitem[{Kingma and Ba(2014)}]{kingma2014adam}
\bibinfo{author}{Kingma, D.P.}, \bibinfo{author}{Ba, J.}, \bibinfo{year}{2014}.
\newblock \bibinfo{title}{Adam: A method for stochastic optimization}.
\newblock \bibinfo{journal}{arXiv preprint arXiv:1412.6980} .
\bibitem[{Li et~al.(2020)Li, Cheng, Wang, Liu and Chen}]{li2020learning}
\bibinfo{author}{Li, S.}, \bibinfo{author}{Cheng, Y.}, \bibinfo{author}{Wang,
  W.}, \bibinfo{author}{Liu, Y.}, \bibinfo{author}{Chen, T.},
  \bibinfo{year}{2020}.
\newblock \bibinfo{title}{Learning to detect malicious clients for robust
  federated learning}.
\newblock \bibinfo{journal}{arXiv preprint arXiv:2002.00211} .
\bibitem[{Ludwig et~al.(2020)Ludwig, Baracaldo, Thomas, Zhou, Anwar, Rajamoni,
  Ong, Radhakrishnan, Verma, Sinn et~al.}]{ibmfl2020ibm}
\bibinfo{author}{Ludwig, H.}, \bibinfo{author}{Baracaldo, N.},
  \bibinfo{author}{Thomas, G.}, \bibinfo{author}{Zhou, Y.},
  \bibinfo{author}{Anwar, A.}, \bibinfo{author}{Rajamoni, S.},
  \bibinfo{author}{Ong, Y.}, \bibinfo{author}{Radhakrishnan, J.},
  \bibinfo{author}{Verma, A.}, \bibinfo{author}{Sinn, M.}, et~al.,
  \bibinfo{year}{2020}.
\newblock \bibinfo{title}{Ibm federated learning: an enterprise framework white
  paper v0. 1}.
\newblock \bibinfo{journal}{arXiv preprint arXiv:2007.10987} .
\bibitem[{McMahan et~al.(2017)McMahan, Moore, Ramage, Hampson and
  y~Arcas}]{mcmahan2017communicationefficient}
\bibinfo{author}{McMahan, H.B.}, \bibinfo{author}{Moore, E.},
  \bibinfo{author}{Ramage, D.}, \bibinfo{author}{Hampson, S.},
  \bibinfo{author}{y~Arcas, B.A.}, \bibinfo{year}{2017}.
\newblock \bibinfo{title}{Communication-efficient learning of deep networks
  from decentralized data}.
\newblock \href{http://arxiv.org/abs/1602.05629}{\tt arXiv:1602.05629}.
\bibitem[{Meidan et~al.(2017)Meidan, Bohadana, Shabtai, Guarnizo, Ochoa,
  Tippenhauer and Elovici}]{meidan2017profiliot}
\bibinfo{author}{Meidan, Y.}, \bibinfo{author}{Bohadana, M.},
  \bibinfo{author}{Shabtai, A.}, \bibinfo{author}{Guarnizo, J.D.},
  \bibinfo{author}{Ochoa, M.}, \bibinfo{author}{Tippenhauer, N.O.},
  \bibinfo{author}{Elovici, Y.}, \bibinfo{year}{2017}.
\newblock \bibinfo{title}{Profiliot: A machine learning approach for iot device
  identification based on network traffic analysis}, in:
  \bibinfo{booktitle}{Proceedings of the symposium on applied computing}, pp.
  \bibinfo{pages}{506--509}.
\bibitem[{Mun and Lee(2021)}]{mun2021internet}
\bibinfo{author}{Mun, H.}, \bibinfo{author}{Lee, Y.}, \bibinfo{year}{2021}.
\newblock \bibinfo{title}{Internet traffic classification with federated
  learning}.
\newblock \bibinfo{journal}{Electronics} \bibinfo{volume}{10},
  \bibinfo{pages}{27}.
\bibitem[{Negka et~al.(2019)Negka, Gketsios, Anagnostopoulos, Spathoulas,
  Kakarountas and Katzenbeisser}]{negka2019employing}
\bibinfo{author}{Negka, L.}, \bibinfo{author}{Gketsios, G.},
  \bibinfo{author}{Anagnostopoulos, N.A.}, \bibinfo{author}{Spathoulas, G.},
  \bibinfo{author}{Kakarountas, A.}, \bibinfo{author}{Katzenbeisser, S.},
  \bibinfo{year}{2019}.
\newblock \bibinfo{title}{Employing blockchain and physical unclonable
  functions for counterfeit iot devices detection}, in:
  \bibinfo{booktitle}{Proceedings of the International Conference on Omni-Layer
  Intelligent Systems}, pp. \bibinfo{pages}{172--178}.
\bibitem[{{Nguyen} et~al.(2019){Nguyen}, {Marchal}, {Miettinen}, {Fereidooni},
  {Asokan} and {Sadeghi}}]{Nguyen2019DioT}
\bibinfo{author}{{Nguyen}, T.D.}, \bibinfo{author}{{Marchal}, S.},
  \bibinfo{author}{{Miettinen}, M.}, \bibinfo{author}{{Fereidooni}, H.},
  \bibinfo{author}{{Asokan}, N.}, \bibinfo{author}{{Sadeghi}, A.},
  \bibinfo{year}{2019}.
\newblock \bibinfo{title}{DÏot: A federated self-learning anomaly detection
  system for {IoT}}, in: \bibinfo{booktitle}{39th IEEE International Conference
  on Distributed Computing Systems}, pp. \bibinfo{pages}{756--767}.
\newblock \DOIprefix\doi{10.1109/ICDCS.2019.00080}.
\bibitem[{Ni{\v{z}}eti{\'c} et~al.(2020)Ni{\v{z}}eti{\'c}, {\v{S}}oli{\'c},
  Gonz{\'a}lez-de, Patrono et~al.}]{nivzetic2020internet}
\bibinfo{author}{Ni{\v{z}}eti{\'c}, S.}, \bibinfo{author}{{\v{S}}oli{\'c}, P.},
  \bibinfo{author}{Gonz{\'a}lez-de, D.L.d.I.}, \bibinfo{author}{Patrono, L.},
  et~al., \bibinfo{year}{2020}.
\newblock \bibinfo{title}{Internet of things (iot): Opportunities, issues and
  challenges towards a smart and sustainable future}.
\newblock \bibinfo{journal}{Journal of Cleaner Production}
  \bibinfo{volume}{274}, \bibinfo{pages}{122877}.
\bibitem[{Rey et~al.(2021)Rey, S{\'a}nchez, Celdr{\'a}n, Bovet and
  Jaggi}]{rey2021federated}
\bibinfo{author}{Rey, V.}, \bibinfo{author}{S{\'a}nchez, P.M.S.},
  \bibinfo{author}{Celdr{\'a}n, A.H.}, \bibinfo{author}{Bovet, G.},
  \bibinfo{author}{Jaggi, M.}, \bibinfo{year}{2021}.
\newblock \bibinfo{title}{Federated learning for malware detection in iot
  devices}.
\newblock \bibinfo{journal}{arXiv preprint arXiv:2104.09994} .
\bibitem[{Sanchez-Rola et~al.(2018)Sanchez-Rola, Santos and
  Balzarotti}]{sanchez2018clock}
\bibinfo{author}{Sanchez-Rola, I.}, \bibinfo{author}{Santos, I.},
  \bibinfo{author}{Balzarotti, D.}, \bibinfo{year}{2018}.
\newblock \bibinfo{title}{Clock around the clock: Time-based device
  fingerprinting}, in: \bibinfo{booktitle}{2018 ACM SIGSAC Conference on
  Computer and Communications Security}, pp. \bibinfo{pages}{1502--1514}.
\bibitem[{S\'anchez~S\'anchez et~al.(2021)S\'anchez~S\'anchez, Jorquera~Valero,
  Huertas~Celdr\'an, Bovet, Gil~P\'erez and
  Mart\'inez~P\'erez}]{sanchez2020survey}
\bibinfo{author}{S\'anchez~S\'anchez, P.M.}, \bibinfo{author}{Jorquera~Valero,
  J.M.}, \bibinfo{author}{Huertas~Celdr\'an, A.}, \bibinfo{author}{Bovet, G.},
  \bibinfo{author}{Gil~P\'erez, M.}, \bibinfo{author}{Mart\'inez~P\'erez, G.},
  \bibinfo{year}{2021}.
\newblock \bibinfo{title}{A survey on device behavior fingerprinting: Data
  sources, techniques, application scenarios, and datasets}.
\newblock \bibinfo{journal}{IEEE Communications Surveys Tutorials}
  \bibinfo{volume}{23}, \bibinfo{pages}{1048--1077}.
\newblock \DOIprefix\doi{10.1109/COMST.2021.3064259}.
\bibitem[{{Thangavelu} et~al.(2019){Thangavelu}, {Divakaran}, {Sairam},
  {Bhunia} and {Gurusamy}}]{Thangavelu2018DEFT}
\bibinfo{author}{{Thangavelu}, V.}, \bibinfo{author}{{Divakaran}, D.M.},
  \bibinfo{author}{{Sairam}, R.}, \bibinfo{author}{{Bhunia}, S.S.},
  \bibinfo{author}{{Gurusamy}, M.}, \bibinfo{year}{2019}.
\newblock \bibinfo{title}{{DEFT}: A distributed {IoT} fingerprinting
  technique}.
\newblock \bibinfo{journal}{IEEE Internet of Things Journal}
  \bibinfo{volume}{6}, \bibinfo{pages}{940--952}.
\bibitem[{Varghese et~al.(2021)Varghese, Wang, Bermbach, Hong, Lara, Shi and
  Stewart}]{varghese2021benchmarking}
\bibinfo{author}{Varghese, B.}, \bibinfo{author}{Wang, N.},
  \bibinfo{author}{Bermbach, D.}, \bibinfo{author}{Hong, C.H.},
  \bibinfo{author}{Lara, E.D.}, \bibinfo{author}{Shi, W.},
  \bibinfo{author}{Stewart, C.}, \bibinfo{year}{2021}.
\newblock \bibinfo{title}{A survey on edge performance benchmarking}.
\newblock \bibinfo{journal}{ACM Comput. Surv.} \bibinfo{volume}{54}.
\newblock \URLprefix \url{https://doi.org/10.1145/3444692},
  \DOIprefix\doi{10.1145/3444692}.
\bibitem[{Xie et~al.(2019)Xie, Koyejo and Gupta}]{xie2019zeno}
\bibinfo{author}{Xie, C.}, \bibinfo{author}{Koyejo, S.},
  \bibinfo{author}{Gupta, I.}, \bibinfo{year}{2019}.
\newblock \bibinfo{title}{Zeno: Distributed stochastic gradient descent with
  suspicion-based fault-tolerance}, in: \bibinfo{booktitle}{International
  Conference on Machine Learning}, \bibinfo{organization}{PMLR}. pp.
  \bibinfo{pages}{6893--6901}.
\bibitem[{Yang et~al.(2019)Yang, Liu, Chen and Tong}]{yang2019federated}
\bibinfo{author}{Yang, Q.}, \bibinfo{author}{Liu, Y.}, \bibinfo{author}{Chen,
  T.}, \bibinfo{author}{Tong, Y.}, \bibinfo{year}{2019}.
\newblock \bibinfo{title}{Federated machine learning: Concept and
  applications}.
\newblock \bibinfo{journal}{ACM Transactions on Intelligent Systems and
  Technology} \bibinfo{volume}{10}, \bibinfo{pages}{1--19}.
\bibitem[{Yin et~al.(2018)Yin, Chen, Kannan and Bartlett}]{yin2018byzantine}
\bibinfo{author}{Yin, D.}, \bibinfo{author}{Chen, Y.}, \bibinfo{author}{Kannan,
  R.}, \bibinfo{author}{Bartlett, P.}, \bibinfo{year}{2018}.
\newblock \bibinfo{title}{Byzantine-robust distributed learning: Towards
  optimal statistical rates}, in: \bibinfo{booktitle}{International Conference
  on Machine Learning}, \bibinfo{organization}{PMLR}. pp.
  \bibinfo{pages}{5650--5659}.

\end{thebibliography}

\end{document}